\documentclass[12pt,letterpaper]{article}
\usepackage[usenames, dvipsnames]{xcolor}
\usepackage{jcapmod}

\usepackage{tocloft}
\usepackage[]{todonotes}
\usepackage{verbatim}
\usepackage{mathrsfs}

\usepackage{bbm} 					
\usepackage{slashed} 				
\usepackage{graphicx}				
\usepackage{subcaption}			
\usepackage{psfrag}				
\usepackage{tensor}				
\usepackage{fouridx}				
\usepackage{bm}					
\usepackage{mdframed}				
\usepackage{multirow}				
\usepackage{soul}					
\usepackage{bbold}				
\usepackage{multicol}				

\usepackage{amsmath}
\usepackage{amssymb}

\usepackage{feynmf}
\usepackage{marvosym}

\usepackage{import}

\newcommand*\xoverline[2][0.75]{%
    \sbox{\myboxA}{$\m@th#2$}%
    \setbox\myboxB\null
    \ht\myboxB=\ht\myboxA%
    \dp\myboxB=\dp\myboxA%
    \wd\myboxB=#1\wd\myboxA
    \sbox\myboxB{$\m@th\overline{\copy\myboxB}$}
    \setlength\mylenA{\the\wd\myboxA}
    \addtolength\mylenA{-\the\wd\myboxB}%
    \ifdim\wd\myboxB<\wd\myboxA%
       \rlap{\hskip 0.5\mylenA\usebox\myboxB}{\usebox\myboxA}%
    \else
        \hskip -0.5\mylenA\rlap{\usebox\myboxA}{\hskip 0.5\mylenA\usebox\myboxB}%
    \fi}
\makeatother

\definecolor{cobalt}{RGB}{44, 98, 120}
\definecolor{celadon}{rgb}{0.67, 0.88, 0.69}
\definecolor{dm}{cmyk}{.20, 0, .30, 0}
\definecolor{burgundy}{rgb}{0.5, 0.0, 0.13}
\definecolor{plotBlue}{RGB}{94, 130, 181}

\newcommand{\lab}[1]{{\mathrm{#1}}}

\newcommand{\M}{M_\lab{pl}}

\hypersetup{
  colorlinks,
  citecolor=Violet,
  linkcolor=cobalt,
  urlcolor=Blue}

\newif\iffastcompile

\fastcompilefalse

\iffastcompile
\newcommand{\js}[1]{}
\newcommand{\jsi}[1]{}
\newcommand{\cl}[1]{}
\newcommand{\lm}[1]{}
\else
\newcommand{\js}[1]{\todo[color=cobalt!30,size=\scriptsize, bordercolor=cobalt!30]{JS: #1}}
\newcommand{\jsi}[1]{\todo[color=cobalt!30,size=\scriptsize, bordercolor=cobalt!30, inline]{JS: #1}}
\newcommand{\cl}[1]{\todo[color=burgundy!30, size=\scriptsize, bordercolor=burgundy!30]{CL: #1}}
\newcommand{\lm}[1]{\todo[color=dm!90, size=\scriptsize, bordercolor=dm!90]{LM: #1}}
\fi

\newcommand{\email}[1]{\href{mailto:#1}{#1}}
\newcommand{\minus}{{\scalebox {0.75}[1.0]{$-$}}}
\newcommand{\dhat}{{\smash{\hat{D}}}}
\ProvideTextCommandDefault{\Dbar}{%
\leavevmode\lower.5ex\rlap{\hskip-.07em\accent"16}D%
}
\newcommand{\Eth}{\text{\Dbar}}

\begin{document}
	\newcommand{\main}{.}
\begin{titlepage}

\setcounter{page}{1} \baselineskip=15.5pt \thispagestyle{empty}

\bigskip\

\vspace{2cm}
\begin{center}
{\fontsize{22}{28} \bfseries Systematics of Axion Inflation\\ \vspace{0.2cm} in Calabi-Yau Hypersurfaces}
\end{center}

\vspace{0.25cm}

\begin{center}
\scalebox{0.95}[0.95]{{\fontsize{14}{30}\selectfont Cody Long, Liam McAllister, and John Stout}}
\end{center}

\begin{center}

\textsl{ Department of Physics, Cornell University, Ithaca, NY 14853, USA}\\

\vspace{0.25cm}
\email{cel89@cornell.edu}, \email{mcallister@cornell.edu}, \email{jes554@cornell.edu }
\end{center}

\vspace{1cm}
{
\noindent We initiate a comprehensive survey of axion inflation in compactifications of type IIB string theory on Calabi-Yau hypersurfaces in toric varieties.
For every threefold with $h^{1,1} \le 4$ in the Kreuzer-Skarke database, we compute the metric on K\"ahler moduli space, as well as the matrix of four-form axion charges of Euclidean D3-branes on rigid divisors.
These charges encode the possibility of enlarging the field range via alignment.  We then determine an upper bound on the inflationary field range $\Delta \phi$ that results from the leading instanton potential, in the absence of monodromy.  The bound on the field range in this ensemble is $\Delta \phi \lesssim 0.3 M_{\rm{pl}}$, in a compactification where the smallest curve volume is $(2\pi)^2\alpha'$, and we argue that the sigma model expansion is adequately controlled.
The largest increase resulting from alignment is a factor $\approx 2.6$.  We also examine a set of threefolds with $h^{1,1}$ up to $100$ and characterize their axion charge matrices.
We discuss how our findings could be modified by the effects of orientifolding, seven-branes, and fluxes.

\noindent
}
 \vspace{0.1cm}

\vspace{0.6cm}
\noindent \today

\end{titlepage}

		\tableofcontents%
\newpage
\section{Introduction}\label{intro}

The prospect of detecting or strongly bounding primordial gravitational waves through measurements of CMB B-modes in the next few years makes the question of large-field inflation in quantum gravity an urgent one.
Exhibiting a totally explicit model of large-field inflation in string theory, or proving no-go theorems that exclude classes of constructions, remains challenging.  A persistent difficulty is establishing control of the theory in the parameter range where large-field inflation would occur: making the inflaton potential flat over a super-Planckian distance often requires adjusting compactification parameters, such as cycle sizes, flux quanta, and numbers of D-branes, away from the weakly coupled limit.  While it is easy to speculate that something that appears difficult might in fact be impossible, and some authors have promoted this expectation to a principle, there has been little success in actually establishing that large-field inflation is (im)possible in some corner of string theory, except in very simple settings.\footnote{See \cite{Baumann:2014nda} for an overview.}

Axion inflation is a promising framework for examining large-field inflation in string theory.  As in the original model of natural inflation \cite{Freese:1990rb}, all-orders shift symmetries give structure to the inflaton potential and sharpen the problem of exhibiting a flat potential over a large range to that of achieving a large axion periodicity.  Axions are numerous in Calabi-Yau compactifications of string theory, descending from $p$-form fields in ten dimensions, reduced on suitable $p$-cycles.
The resulting axion fields inherit perturbatively exact continuous shift symmetries from the higher-dimensional gauge symmetry, provided that the latter is not broken by classical sources such as wrapped D-branes or background fluxes, which would introduce monodromy in the axion potential \cite{SW,MSW}.\footnote{See e.g.~the discussions in \cite{Marchesano:2014mla,McAllister:2014mpa}.}  In this work we will consider axion inflation without explicit monodromy: we will investigate inflation driven by the strictly periodic potential generated by Euclidean D-branes.

Although it is difficult to arrange for a single axion in string theory to have periodicity $2\pi f > M_{\rm{pl}}$
in a regime of perturbative control \cite{Banks:2003sx,Svrcek:2006yi}, an appealing alternative is to arrange for a particular linear combination of $N>1$ axions to have a large effective periodicity.  The resulting inflationary model, aligned natural inflation, is a version of assisted inflation \cite{Liddle:1998jc}. The first such proposal, for the case $N=2$, is due to Kim, Nilles, and Peloso (KNP) \cite{Kim:2004rp}, and is known as `KNP alignment' or `lattice alignment.'
  
More recently, generalizations of lattice alignment to $N \gg 1$ have been studied \cite{Choi:2014rja, Higaki:2014pja, Kaplan:2015fuy}, and a distinct alignment phenomenon involving the kinetic term, known as `kinetic alignment,' has been identified \cite{Bachlechner:2014hsa}.  Related works include \cite{Czerny:2014xja,Tye:2014tja,Kappl:2014lra,Ben-Dayan:2014zsa,Gao:2014uha,Ben-Dayan:2014lca,HT2,Abe:2014pwa,Ali:2014mra,Burgess:2014oma,Garcia-Etxebarria:2014wla,Shiu:2015xda,Palti:2015xra,Kappl:2015esy}.
In \S2 we will review these alignment effects in more detail.  One key point is that the field range enhancement due to lattice alignment is determined by a matrix $Q$ of quantized axion charges carried by instantons, which without loss of generality we can take to be integers.  In an effective field theory construction of aligned natural inflation, the axion periodicity
can be made arbitrarily large if these integer charges are unbounded.  However, quantum gravity theories with conventional black hole thermodynamics are generally thought not to allow exact continuous global internal symmetries.  More concretely, any finite class of string compactifications will be characterized by a finite set of integer data---such as intersection numbers, flux quanta, and D-brane charges---which only allows for a finite degree of alignment.  While this plausibly excludes {\it{arbitrarily}} super-Planckian field ranges in axion theories without monodromy, the question of physical interest is whether the field range $\Delta\phi_{\rm thy}$ allowed by quantum gravity can exceed the upper bound\footnote{For single-field natural inflation, the Planck measurements of the tilt also imply a lower bound on $\Delta\phi$ \cite{Ade:2015lrj}.} $\Delta\phi_{\rm exp}$ determined by measurements of CMB B-modes.

To determine what quantitative upper bound quantum gravity, and in particular string theory, imposes on the field range in axion inflation, one can ask whether the integer data in an actual string compactification can permit a high degree of alignment, and whether this is sufficient to achieve $\Delta\phi_\lab{thy} > \Delta\phi_{\rm exp}$ in a parametrically controlled construction.
In this paper, we answer these questions, in the negative, for a large class of explicit Calabi-Yau compactifications.

We consider inflation driven by the Ramond-Ramond four-form $C_4$, in compactifications of type IIB string theory on Calabi-Yau threefold hypersurfaces in toric fourfolds.  We examine all 5922 threefolds
with $h^{1,1}\le 4$ in the Kreuzer-Skarke database \cite{Kreuzer:2000xy}, and identify divisors that are rigid and so support Euclidean D3-brane contributions to the superpotential.\footnote{Our method is applicable for larger $h^{1,1}$, as we show in \S\ref{largeHodge}, but computing the divisors' topology becomes more expensive.}
In 4390 of these compactifications, Euclidean D3-branes wrapping linear combinations of up to three toric divisors suffice to break all continuous axion shift symmetries, and correspondingly lift all flat directions in the K\"ahler moduli space.\footnote{The flat directions in the remaining examples may well be lifted by more complicated instanton configurations, but we do not analyze those geometries any further.}  The axion fundamental domain is therefore compact in these examples, and we compute its diameter as a function of the K\"ahler moduli.
The geometric field range $\mathcal{R} \approx \Delta\phi_\lab{thy}$, defined in \S\ref{sec:effLag},
is a function
of the curve volume parameters $t_i$, and is homogeneous of degree $-2$ with respect to the overall scaling $t_i \to \lambda t_i$, so the upper bound on $\Delta \phi$ is dictated, in part, by the smallest curve volumes compatible with control of the $\alpha'$ expansion. We argue that in a region of reasonable perturbative control, where the minimum curve volume is $\ell_s^2 \equiv (2\pi)^2\alpha'$, the upper bound on the geometric field range is $\mathcal{R} \lesssim 0.3 \,M_\lab{pl} $, with $M_\lab{pl}$ the four-dimensional reduced Planck mass.

The largest contribution of lattice alignment to $\mathcal{R}$ in our ensemble is a factor of $2.6$, in a compactification where $h^{1,1}=4$ with axion charge matrix
\begin{equation}
\bm{\mathcal{Q}} = 2\pi \begin{pmatrix} ~~ 1 ~ & ~ 0 ~ & ~0 ~ & ~ 0~\\ ~\minus 1 ~ & \minus 1 ~ & ~1 ~ & ~1~ \\
~~0 ~ & ~0 ~ & ~1 ~ & ~0~ \\ ~~0 ~ & ~0 ~ & ~ 0 ~ & ~1~ \end{pmatrix}.
\end{equation}
In this example $\mathcal{R}=0.08 \M$, while with $\mathcal{Q}=2\pi \mathbbm{1}$ one would have $\mathcal{R}=0.03 \M$.

We make one simplifying assumption that deserves special mention.  In determining which divisors $D$ yield Euclidean D3-brane contributions to the superpotential, we examine only the topology of $D$ itself, and require the rigidity condition $h^{\bullet}(D,{\cal{O}}_D)=(1,0,0)$.  We do not systematically include corrections to this zero-mode counting due to orientifolding, worldvolume flux, bulk flux, or intersections with seven-branes (see e.g.~\cite{Blumenhagen:2007bn,Donagi:2010pd,Grimm:2011dj,Bianchi:2011qh,Cvetic:2012ts,Bianchi:2012kt}).
While incorporating these effects is beyond the scope of this work, it will be an important next step.

The organization of this paper is as follows.  In \S\ref{sec1} we review how the topological and geometric data of an O3/O7 orientifold compactification determines an effective theory for axions, and we explain how to compute the field range, including the effects of alignment, in such a theory.  In \S\ref{toric} we recall how to obtain the topological data of a Calabi-Yau threefold hypersurface in a toric variety.  In \S\ref{scan} we present the results of a complete scan through the Kreuzer-Skarke database at $h^{1,1}\le 4$, and in \S\ref{largeHodge} we describe a few examples at much larger $h^{1,1}$.  Our conclusions appear in \S\ref{sec:conclusions}.

\section{Four-Form Axions in O3/O7 Orientifolds} \label{sec1}

A comparatively well-understood class of four-dimensional $\mathcal{N} = 1$ solutions of string theory are compactifications of type IIB string theory on O3/O7 orientifolds of Calabi-Yau threefolds.  Because the full space of $\mathcal{N} = 1$ orientifolds is not known,\footnote{However, see \cite{Gao:2013pra} for progress in classifying involutions that exchange two coordinates.} in this work we will focus on their Calabi-Yau double covers, which can be enumerated systematically in the case of hypersurfaces in toric varieties.

\subsection{The effective Lagrangian} \label{sec:effLag}
		In type IIB string theory compactified on an O3/O7 orientifold of a Calabi-Yau threefold $X$, the closed string moduli are the complex structure moduli, axiodilaton, and K\"ahler moduli. The complex structure moduli and axiodilaton can be completely fixed by a suitable choice of quantized $G_3$ flux, while the K\"ahler moduli are unfixed to all orders in perturbation theory due to the gauge symmetry of the Ramond-Ramond four-form.
When $h^{1,1}_{-} = 0$, which we will assume in this work, the coordinates on K\"ahler moduli space are the complexified volumes $T^i$ of four-cycles, defined as
		\begin{equation}
			T^i = \frac{1}{2}\int_{D^i} \! J\wedge J + i\int_{D^i}\! C_{4} \equiv \tau^i + i \theta^i,
		\end{equation}
where $J$ is the K\"{a}hler form, $D^i$ is a basis element of $H_4(X,\mathbbm{Z})$, and $C_{4}$ is the Ramond-Ramond four-form field. The K\"{a}hler potential is given by
\begin{equation}\label{kclass}
\mathcal{K} = -2\,\log\mathcal{V},
\end{equation}
where $\mathcal{V}$ is the volume\footnote{All volumes in this work are determined in ten-dimensional Einstein frame in units of $\ell_s = 2\pi\sqrt{\alpha'}$.}
of the internal space,
\begin{equation}
\mathcal{V} = \frac{1}{6}\int_X \! J\wedge J\wedge J.
\end{equation}
We can write the volume as $\mathcal{V} = \frac{1}{6} \kappa^{ijk} t_i t_j t_k$ by expanding the K\"{a}hler form as $J = t_i \omega^i$, where $\omega^i$ form a basis for $H^{1,1}(X, \mathbbm{Z})$ and the $\kappa^{ijk}$ are triple intersection numbers among divisors $D^i$.

The space of K\"ahler parameters $t_i$ is restricted by the requirement that the metric on field space be positive definite.  To identify the resulting conditions on the $t_i$, we consider the Mori cone of $X$, ${\rm{Mori}}(X)$, which is the cone of holomorphic curves: any holomorphic curve $C$ in $X$ can be written as
\begin{equation}
C=\sum_a n_a C_a
\end{equation}
where the $C_a$ are the generators of ${\rm{Mori}}(X)$, and $n_a$ are nonnegative integers.
The K\"ahler cone is the space dual to the Mori cone, i.e.~it is the region of K\"ahler parameters $t_i$ for which $\int_C J > 0$ for every holomorphic curve $C$.

Everywhere inside the K\"ahler cone, the axion field space metric $K_{ij}$ obtained from the tree-level K\"ahler potential (\ref{kclass}) is positive definite.  However, as one approaches the walls of the K\"ahler cone, (\ref{kclass}) does not necessarily provide a good approximation to the true K\"ahler potential that incorporates all $\alpha'$ and $g_s$ corrections.  Our computation based on (\ref{kclass}) is therefore meaningful only when the $t_i$ are restricted to a proper subset of the K\"ahler cone.  To understand the conditions that must be imposed on the $t_i$, we recall the form of the perturbative and nonperturbative corrections to the effective Lagrangian.  The superpotential for the K\"ahler moduli is purely nonperturbative because of the axion shift symmetry, and we will compute it directly in this work, modulo some important technical assumptions detailed below.  The K\"{a}hler potential receives perturbative corrections in the $\alpha'$ and $g_s$ expansions, as well as nonperturbative corrections, and none of these has been fully characterized.

Control of the string loop expansion can be achieved by arranging for $g_s \ll 1$ by a suitable choice of quantized three-form flux.  We remark that string loop corrections to the K\"{a}hler potential are suppressed not only by powers of $g_s$, but also by powers of ${\cal V}$, so at large threefold volume very small $g_s$ is not necessary for ensuring that string loop corrections are small.\footnote{Investigations of axion field ranges at moderately strong coupling include  \cite{Grimm:2014vva,Conlon:2016aea}.}  Next, as a proxy for control of the $\alpha'$ expansion, we will consider worldsheet instantons wrapping nontrivial curves $C \subset X$: in the region where the $g_s$ and $\alpha'$ expansions are well-controlled, these are generically the leading nonperturbative corrections to $K$, and are proportional to\footnote{We adopt the normalizations of \cite{Denef:2005mm}, as laid out in Appendix A of \cite{Denef:2005mm}.}  
\begin{equation}
	\Delta \mathcal{K} \sim {\cal{V}}^{-1}\,e^{-2 \pi \sqrt{g_s} \,t},
\end{equation}
where $t$ is the Einstein frame volume of $C$, i.e.~the volume measured with the ten-dimensional Einstein frame metric, in units of $\ell_s^2$.  (The string frame volume of $C$ is then $\sqrt{g_s}t$.)  To ensure that the worldsheet instanton corrections are small, we will require that the volumes of all curves are larger than some threshold value.  In this work we take the threshold volume to be $\ell_s^2$, so that worldsheet instanton contributions are suppressed by factors of $e^{-2\pi\sqrt{g_s}}$, which is small for $g_s \gtrsim 0.1$.
Because the K\"ahler metric is homogeneous of degree $\minus 2$ with respect to overall scaling $t_i \to \lambda t_i$, it is trivial to translate our results to any other desired threshold, as might be motivated by examining the form of perturbative corrections in particular examples.

In view of the above requirement, we now define
the \emph{stretched K\"{a}hler cone} as the set of K\"{a}hler parameters $t_i$ for which $\int_C J > 1$, for all holomorphic curves $C$.  The condition on curve volumes explained in the previous paragraph corresponds to the requirement that the $t_i$ lie in the stretched K\"{a}hler cone.  This condition leads to a lower bound on the volumes of divisors, $\tau^i \equiv \partial \mathcal{V}/\partial t_i$, and on the volume $\mathcal{V}$ of $X$ itself.

\subsection{The axion fundamental domain} \label{sec:axionfund}

At a point in K\"ahler moduli space that falls inside the stretched K\"ahler cone, the effective Lagrangian for the $N = h^{1,1}$ axions
takes the form, in four-dimensional Einstein frame,
\begin{equation}
\mathcal{L} = \frac{M_\lab{pl}^2}{2}{\cal{R}}_4-\frac{M_\lab{pl}^2}{2}K_{ij} \partial^\mu \theta^i \partial_\mu \theta^j - \sum\limits_{a=1}^P \Lambda_a^4 \big(1 - \cos(Q^a_{\ i} \theta^i)\big).
\label{eq:eft}
\end{equation}
Here $K_{ij}$ is the K\"ahler metric on field space, and $(2\pi)^{-1}Q$ is a matrix of rational numbers determined by instanton charges. We will search for examples in which $Q$ is a full-rank (that is, rank $N$) matrix, so that there are no exactly flat directions in the axion field space, and correspondingly no unstabilized\footnote{Strictly speaking, we will not be stabilizing the real part K\"ahler moduli $\tau_i$, in the sense that we will not minimize the scalar potential with respect to the $\tau_i$.  We do, however, ensure that all the $\tau_i$ appear in the superpotential, in $N$ linearly independent combinations.} K\"ahler moduli.  In order for $Q$ to have rank $N$, there must be at least $N$ linearly independent divisors contributing to the superpotential, i.e.~we must have $P \geq N$.

\begin{figure}
	\centering
	\includegraphics[width=0.64\textwidth]{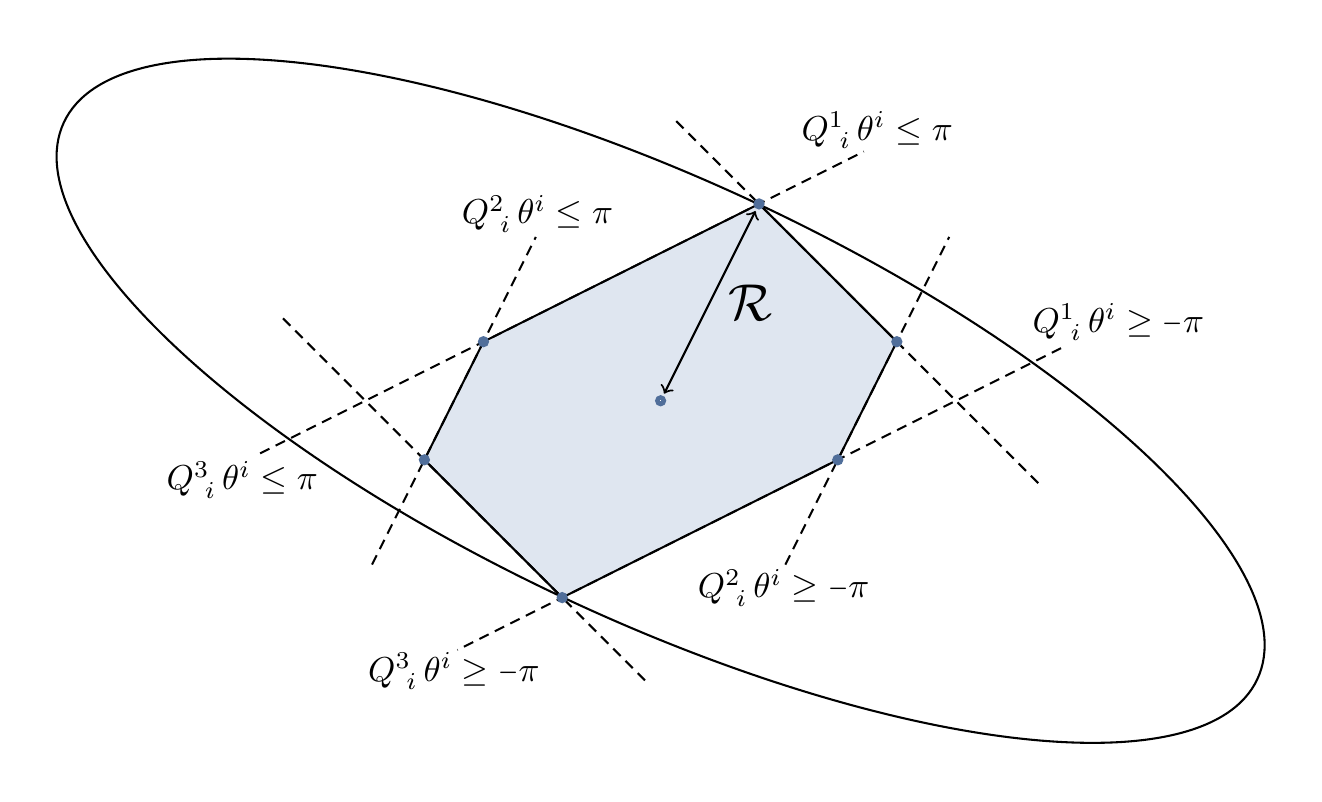}
	\caption{The geometric field range $\mathcal{R}$ is the semi-diameter of the fundamental domain $\mathcal{F}$, which is the region contained in the intersection of the $2P$ hyperplane constraints $-\pi \leq Q\indices{^a_i}\theta^i \leq \pi$. Surfaces of constant distance are ellipsoids with weight matrix $K_{ij}$.}
	\label{fig:fundDomain}
\end{figure}

The \emph{fundamental domain} $\cal{F}$ (cf.~\cite{Bachlechner:2014gfa}) of the axions is the region contained in the intersection of the $2P$ half-plane constraints $\minus\pi \leq Q\indices{^a_i} \theta^i \leq \pi$, as visualized in Figure \ref{fig:fundDomain}.  When $Q$ has rank $N$, $\cal{F}$ is compact.

The fundamental domain is a polytope in field space, and may also be expressed as the convex hull of a set of vertices $\{\mathbf{d}_i\}$.  We define the \emph{geometric field range} $\mathcal{R}$ as the distance, measured with respect to the K\"{a}hler metric $K_{ij}$, from the origin to the most distant point on the boundary of $\cal{F}$.  Equivalently, $\mathcal{R}$ is the distance from the origin to the most distant of the $\mathbf{d}_i$, i.e.~$\mathcal{R}$ is the semi-diameter of $\cal{F}$.

The length $\Delta \phi$ of an inflationary trajectory driven by a general potential on $\cal{F}$ may be larger or smaller than $\mathcal{R}$, but when the initial conditions are arranged
so that the trajectory is well-approximated by a straight line, we expect that $\Delta \phi \lesssim \mathcal{R}$.
We have verified this expectation by solving for the inflationary evolution that results from the full potential.

The identifications defining $\cal{F}$, and hence also the size $\mathcal{R}$ of $\cal{F}$, depend on the set of instantons included in the sum in (\ref{eq:eft}).
Because the $\Lambda_a$ depend exponentially on four-cycle volumes, there will generally be large hierarchies among the $\Lambda_a$, and so some terms in the axion potential may provide only small ripples that are unimportant in determining the maximum field range. Our approach is to choose the dominant instantons, defined as follows.  Given a set of $P>N$ instanton contributions, i.e.~$P$ row vectors $Q\indices{^1_i},\ldots,Q\indices{^P_i}$, one can search for one or more sets of $N$ linearly-independent vectors, corresponding to
full-rank square matrices contained in $Q$.
When there are multiple such full-rank sets, we choose the one for which the $\Lambda_a$ are as large as possible; that is, we identify the $2N$ most important hyperplanes defining the fundamental domain.\footnote{Specifically, we sort the $P>N$ vectors so that the corresponding $\Lambda_a$ are ordered from largest to smallest.  We then select vectors in order from this list, omitting any vector that is not linearly independent of those that have already been selected, and so arrive at a set of $N$ vectors that can be assembled to form a full-rank square matrix $\mathcal{Q}$.}

Once the dominant rows of $Q$ are identified, the corresponding inequalities
define a polytope in field space.  The point in this convex polytope furthest from the origin must be one of the vertices $\mathbf{d}_i$.
Thus, given a constant K\"{a}hler metric $\mathbf{K}$ and a full-rank square matrix $\mathcal{Q}\indices{^i_j} \subset Q\indices{^a_i}$, corresponding to the identifications imposed by the leading instantons, we obtain the axion field range by enumerating the vertices of the associated polytope and computing
\begin{equation}
	\mathcal{R}^2 = \underset{i}{\lab{max}}\,\, \mathbf{d}_i^\intercal \!\cdot \!\mathbf{K}\!\cdot\! \mathbf{d}_i .
	\label{eq:geomfieldrange}
\end{equation}
Each choice of $\mathcal{Q}$ will determine a different polytope in field space and thus yield a different value of $\mathcal{R}$.  In particular, the semi-diameter of the polytope formed by the $2N$ most important hyperplanes serves as an upper bound on the length of straight-line trajectories that stay within the fundamental domain.

\subsection{The superpotential} \label{wsec}

In the type IIB orientifolds considered in this work, the
superpotential interactions of the K\"{a}hler moduli $T^i$ are generated by nonperturbative effects, either from Euclidean D3-branes on a divisor $D$ in the Calabi-Yau $X$, or from strong gauge dynamics, such as gaugino condensation, on a stack of seven-branes on a divisor $D$ in $X$.  As explained above, we will restrict our attention to Euclidean D3-branes.  Necessary and sufficient conditions for a Euclidean D3-brane contribution
to the superpotential were given in~\cite{Witten:1996bn}.  These conditions were derived in the case of M-theory
compactified on an elliptically fibered Calabi-Yau fourfold $Y_4$, with base $B_3$.
Consider a Euclidean M5-brane wrapping a smooth
divisor $\hat{D} \subset Y_4$.
Two necessary conditions for a superpotential contribution are that $\hat{D}$ is \emph{vertical}, meaning that $\pi(\hat{D})$ is a divisor of $B_3$, and that $\hat{D}$ is \emph{effective} (see e.g.~\cite{nla.cat-vn2659416} for the definition). Granting these requirements, a final condition sufficing for a contribution is the rigidity condition
	\begin{equation}	
		h^{\bullet}(\hat{D}, \mathcal{O}_\dhat) = (1, 0, 0, 0).
		\label{eqn:rigid}
	\end{equation}
We will refer to divisors that obey these conditions as {\emph{rigid divisors}}, with the vertical and effective conditions being implicit.

To translate (\ref{eqn:rigid}) to a condition on smooth
divisors $D=\pi(\hat{D})\subset B_3$, we use the relation~\cite{1997alg.geom..4008G}
\begin{equation}
h^i(\hat{D},\mathcal{O}_\dhat) = h^i(D,\mathcal{O}_D) + h^{i-1}(D,-\Delta|_{D}), \quad 0 \leq i \leq 3,
\end{equation} where $h^{a}\equiv 0$ when $a<0$.
Here $12\Delta = \sum n_i \Sigma^i$, where the $\Sigma^i$ are the loci where the fiber degenerates, and the $n_i$ denote the type of singularity.
Since the $h^{i-1}(D,-\Delta|_{D})$ are nonnegative, a necessary condition on $D$ in order for $\hat{D}$ to fulfill the sufficient condition (\ref{eqn:rigid}) is that $h^i(D,\mathcal{O}_D) = 0$, $i=1,2$.  In the special case that the degeneration locus of the elliptic fiber does not intersect $D$---in weak coupling terms, this means that $D$ does not intersect divisors wrapped by D7-branes---we have $h^{i-1}(D,-\Delta|_{D})=0$, so that
\begin{equation} \label{dcondition}
h^{\bullet}(D,\mathcal{O}_D) = (1,0,0)
\end{equation}
actually suffices to ensure a superpotential contribution.  In summary, a divisor $D\subset B_3$ that is effective, does not intersect the discriminant locus $12\Delta$, and obeys the rigidity condition (\ref{dcondition}) supports a Euclidean D3-brane contribution to the superpotential: its preimage $\hat{D} = \pi^{*}(D)$ is effective, vertical, and obeys (\ref{eqn:rigid}).

We have emphasized the `threefold rigidity condition' (\ref{dcondition}) because it depends only on the base
$B_3$, and so can be assessed directly from the  combinatorial data in the Kreuzer-Skarke database.  A more comprehensive analysis, also applicable to divisors $D$ that intersect $\Delta$, would require information about the elliptic fibration, which in our framework requires specifying an orientifold of the Calabi-Yau threefold $X$ whose image is the non-negatively curved base $B_3$.
A systematic treatment of all $\mathbbm{Z}_2$ involutions is beyond the scope of this work.
We will work with
the K\"ahler potential $\mathcal{K} = -2\log\left(\tfrac{1}{2}\mathcal{V}\right)$, where $\mathcal{V}$ is the volume of the double-cover Calabi-Yau manifold.
This provides a reasonable proxy for the metric on the K\"ahler moduli space of the orientifold, at least in the case of orientifolds that flip a single toric coordinate $x_i \rightarrow -x_i$.  In such a case, we have $h^{1,1}_+ = h^{1,1}(X)$, and we do not expect the orientifold action to significantly change the intersection ring.
Beyond the effects of orientifolds themselves, it is worth noting that incorporating D7-branes provides additional freedom to increase the field range $\mathcal{R}$, by factors of the dual Coxeter numbers of the condensing gauge groups.\footnote{A string theory embedding of this proposal was considered in~\cite{Long:2014dta}, where the enhancement was realized by multiply-wound D7-branes.}

Let us be very clear on this point: an effective divisor $D$ obeying (\ref{dcondition}) that does not intersect seven-branes (including O7-planes) will yield a Euclidean D3-brane superpotential term; but because we are working directly with threefolds, without either orientifolding or taking a weak-coupling limit from a fourfold, the non-intersection condition is a simplifying assumption that is not verifiable in our framework.  We view this approach as an intermediate step between working only with the $\mathcal{N} = 2$ data of a threefold, and performing a full $\mathcal{N} = 1$ analysis complete with explicit orientifolding.

The superpotential that results takes the form
\begin{equation}
W = W_0 + \sum_{\alpha=1}^{p} A_\alpha \,e^{-2\pi q\indices{^\alpha_i} T^i},
\end{equation}
where $W_0$ is a flux-dependent constant, and $A_\alpha$ are Pfaffians that depend on the vacuum expectation values of the complex structure moduli. The constant matrix $q\indices{^\alpha_i}$ specifies which K\"ahler moduli appear in each non-perturbative contribution to the superpotential; at the level of our analysis each of the $p$ linear combinations $\Eth^{\alpha} \equiv q\indices{^\alpha_i} D^i$ corresponds to a rigid divisor. The full supergravity potential is given by
\begin{equation} \label{sugra}
V = e^{\mathcal{K}}\left({\cal{D}}_i W \overline{\mathcal{D}^i W} - 3|W|^2\right),
\end{equation}
where ${\cal{D}}_i = \partial_i + K_{i}$ is the K\"ahler covariant derivative.
In this work, we will assume that the moduli can be stabilized in a vacuum where the cosmological constant is small in string units,
and that the dynamics of the real-part saxions may be ignored.\footnote{Ignoring the saxions would be untenable in a construction of an inflationary solution, but is reasonable here because we are simply deriving upper bounds on the geometric diameter.}
The effective Lagrangian density for the axions $\theta^i$ is then given by
Eq. \ref{eq:eft}, where
\begin{equation} \label{bigQ}
	Q\indices{^a_i} = 2\pi\begin{pmatrix} q\indices{^\alpha_i} \\ q\indices{^\beta_i} - q\indices{^\gamma_i} \end{pmatrix},
\end{equation}
is a $P \times N$ matrix, with $P = p(p+1)/2$.  The last $p(p-1)/2$ rows consist of differences $q\indices{^\beta_i} - q\indices{^\gamma_i}$ with $\beta > \gamma$, which result from cross terms in (\ref{sugra}).

In summary, the axion charge matrix $Q$, whose rows specify the hyperplanes that define the fundamental domain,
is given by (\ref{bigQ}), where $\Eth^{\alpha} = q\indices{^\alpha_i} D^i$, $\alpha=1,\ldots, p$ are $p$ effective divisors of the threefold $X$ that fulfill the rigidity condition (\ref{dcondition}), and so support Euclidean D3-brane contributions to the superpotential.
We now turn to understanding the impact of the axion charge matrix $Q$ on the size ${\cal R}$ of the fundamental domain.

\subsection{Computing the field range} \label{computerange}

In \S\ref{wsec} we explained how to obtain the data of the periodic identifications defining the axion fundamental domain ${\cal F}$, which are determined by the particular divisors $\Eth^{\alpha}$ that are rigid and so support Euclidean D3-brane superpotential terms.  These identifications correspond to the hyperplanes in Figure \ref{fig:fundDomain}.
We also recalled, in \S\ref{sec:effLag}, how to compute the K\"ahler metric $K_{ij}$ and to determine the region of the two-cycle size parameters $t_i$ for which the $\alpha^{\prime}$ and $g_s$ expansions are well-controlled (the `stretched K\"ahler cone').  This metric corresponds to the ellipse in Figure \ref{fig:fundDomain}.
These data completely specify the geometry of ${\cal F}$, or more precisely the possible geometries of ${\cal F}$: the size ${\cal R}$ of ${\cal F}$ depends on the K\"ahler moduli.  To determine the maximal field range in a given theory, we must maximize ${\cal R}$ subject to the linear constraints on the $t_i$ that define the stretched K\"ahler cone.

The tree-level metric is a homogeneous function of the $t_i$, and scales with the overall volume as $\mathcal{V}^{-4/3}$.
By using the scaling $t_i \to \lambda t_i$, one finds that the maximal field range is achieved on the boundary of the stretched K\"ahler cone. If ${\rm{Mori}}(X)$ is simplicial, the $h^{1,1}$ constraints $\int_{C_i}J = 1$ can be simultaneously fulfilled at the apex of the stretched K\"ahler cone, where all of the two-cycle volumes are set to unity (in units of $\ell_s$), and the maximal field range is achieved at the apex. However, in more general cases the point in the stretched K\"ahler cone giving the largest ${\cal R}$ can occur on a wall, but away from the apex.  We therefore searched the stretched K\"ahler cone numerically to determine the optimal field range.
For the purpose of the search, we retained only $h^{1,1}$ terms in the potential, taking
\begin{equation}
V = \sum_{i=1}^{h^{1,1}}\left( 1 - \text{cos}\left( \mathcal{Q}\indices{^{i}_{j}}\theta^j\right)\right),
\label{eq:infPot}
\end{equation}
where $\mathcal{Q}\indices{^i_j}$ is the leading-order\footnote{After the $h^{1,1}$ most important terms have been determined, by comparing their prefactors $\Lambda_a$ according to the algorithm given in \S\ref{sec:axionfund}, the problem becomes purely geometric, and we can then set all $\Lambda_a = 1$, as we have done in (\ref{eq:infPot}).} full rank piece of the full $Q$.  In general there will be further terms that reduce the size the of the fundamental domain, both from additional instantons and from cross terms in the supergravity potential, but because we are quoting an upper bound these can be omitted at this stage.

To search for the maximal ${\cal R}$, we computed the four-cycle volumes at a reference point $\mathbf{t}_0$, and then extracted the full rank piece of $Q$ that is leading order at $\mathbf{t}_0$.  We then scanned over the stretched K\"ahler cone for the point $\mathbf{t}_L$ with the largest $\mathcal{R}$.
For the reference point, we used the apex of the stretched K\"ahler cone, defined as the point where the Euclidean norm of the vector $(v_1,\ldots, v_{N_C})$ is minimized, where the $v_a \equiv \int_{C_a}\!J$ and $C_a$ are the $N_C$ generators of the Mori cone.
We then checked that $\mathcal{Q}$ at $\mathbf{t}_L$ is the same as at $\mathbf{t}_0$, meaning that the same instantons remain dominant, and the analysis is self-consistent.  In a small fraction of cases we found that the set of dominant instantons changed during the exploration from $\mathbf{t}_0$ to $\mathbf{t}_L$, which we then accounted for in computing the field range.

\subsection{Alignment}

Many authors have argued that quantum gravity will censor super-Planckian field displacements, or at least will do so in sufficiently restrictive circumstances.
The large degree of structure imposed on axion theories by all-orders shift symmetries makes these theories a promising setting for directly quantifying the restrictions, if any, that descend from quantum gravity.  The objective of the present work is to compute the size\footnote{We stress that including a source of monodromy, which we will not do here, may ultimately allow displacements ${\cal{O}}(n{\cal R})$, where $n\in \mathbbm{Z}$ is the number of cycles (also known as windings) of monodromy.} ${\cal R}$ of the axion fundamental domain ${\cal F}$ in an ensemble of string compactifications.

Once the rigid divisors $\Eth^{\alpha} \equiv q\indices{^\alpha_i} D^i$, the K\"ahler metric $K_{ij}$, and the stretched K\"ahler cone have been determined in a particular theory, the size ${\cal R}$ of ${\cal F}$ is completely specified, and one could mechanically apply the process described in \S\ref{computerange} to compute ${\cal R}$ in a large number of examples, as we shall do in \S\ref{scan}.  However, it will be valuable to first explain that a suitable structure in the axion charge matrix could lead to ${\cal R} \gg \M$, even while the eigenvalues of $K_{ij}$ remain $\ll \M^2$: this is the celebrated phenomenon of \emph{alignment}, and more precisely of KNP alignment \cite{Kim:2004rp}, also known as lattice alignment.  Here we will attempt to be very precise about the notion of lattice alignment in a Calabi-Yau compactification.

Roughly speaking, an axion theory may be said to manifest lattice alignment when the size of ${\cal F}$ is larger than it `would have been if $Q$ had been trivial,' i.e.~the notion of alignment is that of an increase in field range resulting from the structure of the axion charge matrix $Q$.
Heuristically, one might try to define the alignment enhancement\footnote{When $\eta>1$, we say that the theory manifests alignment, and when $\eta<1$ the result may be termed anti-alignment.} factor $\eta$ as
\begin{equation} \label{naive}
\eta_{\rm{naive}} \overset{?}{=} \frac{\mathcal{R}_\lab{actual}}{\mathcal{R}_{\lab{Q=2\pi\mathbbm{1}}}}\,.
\end{equation}
The numerator is well-defined in general, but the axion charge matrix \emph{alone} is not invariant under a change of the variables $\theta^i$, so stating that $Q = 2 \pi \mathbbm{1}$ presupposes a choice of basis.  The (physically meaningful) field range ${\cal R}$ is of course invariant under the change of variables $\theta^i \to M\indices{^i_j}\theta^j$, with $M \in GL(N,\mathbbm{R})$, but $K_{ij}$ and $Q$ separately transform.

Why talk about alignment at all, if a precise definition is subtle (though achievable, see below)?  One motivation is that it is generally far easier to compute the classical geometric data determining the metric $K_{ij}$ than it is to determine the nonperturbative, quantum data of $Q$, which after all is a matrix of axionic charges carried by (D-brane) instantons.  As such, one may sometimes know $K_{ij}$ without knowing $Q$, and it would then be valuable to understand how large an error might be made by approximating $Q \approx 2 \pi \mathbbm{1}$.  In systems of $N \gg 1$ axions, including the ensemble studied here with $2\le N \le 100$, this error can easily be a factor of order $N$, and in theories with special structure \cite{Choi:2014rja,Kaplan:2015fuy} (not established to date in string theory) the error can be exponential in $N$.

If we were equipped with a canonical choice of basis ${\cal B}$, we could define the denominator in (\ref{naive}) by taking the `reference' charge matrix to read $Q = 2 \pi \mathbbm{1}$ {\it{in the basis}} ${\cal B}$.  In other words, the degree of alignment would be dictated by the extent to which the actual charge matrix $Q$, expressed in the basis ${\cal B}$, differs from $2 \pi \mathbbm{1}$, as quantified by (\ref{naive}).

We are not aware of a natural and fully-specified canonical basis.  However, a natural but (in general) overcomplete set consists of the minimal generators of $\lab{Eff}(X)$, the cone of effective divisors in $X$.  The number $N_{\lab{Eff}}$ of minimal generators ${\cal{E}}_A$, $A=1,\ldots,N_{\lab{Eff}}$ of $\lab{Eff}(X)$ often exceeds $h^{1,1}$, and there is then no unique choice of a basis for $H^{1,1}(X)$: there are finitely many choices.

Assume for the moment that $N_{\lab{Eff}}=h^{1,1}$, so that the generators ${\cal{E}}_A$ of $\lab{Eff}(X)$ define a unique basis ${\cal B}$.  If each of the ${\cal{E}}_A$ were rigid and supported a Euclidean D3-brane contribution to the superpotential, we would have $\mathcal{Q} = 2 \pi \mathbbm{1}$ in the basis ${\cal B}$ defined by the ${\cal{E}}_A$.  Moreover, because every effective divisor is a linear combination of the ${\cal{E}}_A$ with nonnegative integer coefficients, the Euclidean D3-branes supported on the ${\cal{E}}_A$ correspond to the most important instanton contributions in the theory: any additional rigid divisors will have equal or larger action.
This simple theory, in which the minimal generators ${\cal{E}}_A$ of $\lab{Eff}(X)$ are rigid, serves as a reference case that we define to have trivial alignment ($\eta=1$).

We now propose that a natural definition of a trivial charge matrix $Q$ is the matrix whose rows are the minimal generators ${\cal{E}}_A$ of $\lab{Eff}(X)$, even when $N_{\lab{Eff}}>h^{1,1}$.  In other words, a well-defined null hypothesis for examining alignment is the assumption that each of the ${\cal{E}}_A$ gives an independent contribution to the non-perturbative superpotential.
We may then define the enhancement factor $\eta$ as
\begin{equation}
	\eta = \frac{\mathcal{R}_\lab{actual}}{\mathcal{R}_{\lab{Eff}(X)}},
\end{equation}
where $\mathcal{R}_\lab{actual}$ is computed using the $Q$ generated by the rigid divisors $\Eth^{\alpha}$, and $\mathcal{R}_{\lab{Eff}(X)}$ is computed using the (by definition) trivial $Q$ generated by assuming that the minimal effective divisors ${\cal{E}}_A$ are rigid.  For both the numerator and the denominator only the $h^{1,1}$ most important rows of $Q$ are included, as explained in \S\ref{sec:axionfund}.

Although we have now given a precise definition of the enhancement $\eta$ resulting from lattice alignment, it remains to determine whether $\eta$ can be large in actual string compactifications.  We therefore turn to determining the numbers $q\indices{^\alpha_i}$ in an ensemble of Calabi-Yau geometries.

\section[The Topology of Calabi-Yau Hypersurfaces]{The Topology of Calabi-Yau Hypersurfaces}\label{toric}
Calabi-Yau hypersurfaces in toric fourfolds provide a large ensemble of Calabi-Yau threefolds, and allow an efficient combinatorial approach to determining the geometry~\cite{1993alg.geom.10003B, Kreuzer:2000xy}. We refer the reader to~\cite{cox2011toric}, among many others, for an introduction to the subject.

The combinatorial data needed to construct a Calabi-Yau consists of a dual pair of reflexive polytopes $\Delta$ and $\Delta^\circ$, and a triangulation of $\Delta^\circ$ that defines a fan $F$. $F$ then defines a toric variety $V$, and the anticanonical hypersurface $-K$ in $V$ is a Calabi-Yau threefold $X$. The triangulation of $\Delta^\circ$ must be \emph{star} with respect to the origin, meaning that every simplex must contain the origin, in order to define a fan. In addition, the triangulation must be \emph{fine} and \emph{regular}, in order to ensure that the hypersurface is generic and projective.\footnote{See~\cite{Altman:2014bfa} for a discussion of these points.}  Because a generic hypersurface misses any given point of $V$, we can allow $V$ to have pointlike singularities without making a generic threefold singular.  As a result, points interior to facets can be ignored when triangulating $\Delta^\circ$.

We have made use of several  publicly-available software packages to obtain and analyze triangulations.  The algebraic software {\tt{Sage}}~\cite{sage} provides a useful interface for working with toric varieties. The triangulations can be performed in {\tt TOPCOM}~\cite{Rambau:TOPCOM-ICMS:2002}, which has been integrated into {\tt Sage}. In addition, we used the program {\tt PALP}~\cite{Kreuzer:2002uu} for calculations involving reflexive polytopes, and its {\tt Mori} extension~\cite{Braun:2011ik} is very powerful in computing relevant topological data at small $h^{1,1}$.
Most triangulation algorithms are not specialized to compute star triangulations; instead, all triangulations are computed, and then the star ones are selected.  When one is mostly concerned with hypersurfaces with small $h^{1,1}$, whose polytopes are readily triangulated in {\tt TOPCOM}, the cost of computing all triangulations is generally not prohibitive.  However, since we will describe some preliminary results at large $h^{1,1}$, we will outline how one can begin to probe these geometries. For $h^{1,1} \lesssim 30$, one can use the algorithms given in~\cite{Long:2014fba,Altman:2014bfa} to get all the triangulations of the polytope by gluing together the triangulations of individual facets, but this quickly becomes expensive as $h^{1,1}$ grows.
However, even when computing all triangulations in this way is impractical, it is possible to obtain a single triangulation very quickly.  The method was implicit in~\cite{1993alg.geom.10003B}, and was made very clear in~\cite{Braun:2014xka}: one simply computes a regular and fine (not star) triangulation of the polytope, and then deletes the lines in the strict interior of the polytope. This induces a regular triangulation of the facets, and then a star triangulation is constructed by drawing a line from the origin to each point in the polytope.  Using this method it is easy to compute a single triangulation of any polytope in the Kreuzer-Skarke database; for instance, a triangulation of a polytope whose hypersurface $X$ has $h^{1,1}(X) = 400$ takes about ten seconds on a typical laptop.

The tree-level K\"ahler potential depends only on the classical volume, and can be computed easily via toric methods, as one only needs the intersection ring and the Mori cone ${\rm{Mori}}(X)$.  We will consider only \emph{favorable} hypersurfaces $X$, i.e.~those in which all of the divisors of the Calabi-Yau are inherited from divisors of $V$; in such cases we have ${\rm{Mori}}(X) \subset {\rm{Mori}}(V)$.  Computing ${\rm{Mori}}(X)$ from toric data is challenging, so we take the conservative approach of imposing the Mori cone conditions inherited from $V$.\footnote{Note that ${\rm{Mori}}(X)$ can be a proper subset of ${\rm{Mori}}(V)$.  In particular, if a curve $C$ is in $V$ but not in $X$, the sigma model expansion on $X$ is unaffected by taking the volume of $C$ to zero.}

Determining the nonperturbative superpotential is more involved, as we need to know the Hodge numbers of divisors in the hypersurface. In favorable Calabi-Yau threefolds,  the vanishing loci of the individual homogeneous coordinates, corresponding to rays in the fan, furnish a generating set of $h^{1,1} +4$ divisors $\check{D}^a$ in the Calabi-Yau.  To search for a set of $h^{1,1}$ independent rigid divisors we consider the cohomology of these generators and their linear combinations.
Recall that the number of independent homology classes of divisors is counted by $h^{1,1}(X,\mathbbm{Z})$. Given a choice of a basis $\{D^i\}$ of divisors, the task at hand is to determine whether a divisor $D = \sum_i a_i D^i$ is rigid. To do so, we need to specify what values the $a_i$ can take. In some cases one can choose a basis such that all holomorphic hypersurfaces can be written as sums of the $D^i$ with non-negative integer coefficients, and the problem reduces to scanning over an $(\mathbbm{N})^{h^{1,1}}$ lattice. This happens only when $\lab{Eff}(X)$ is simplicial.\footnote{The divisors whose rigidity properties we need to examine are all the divisors $D$ that are effective in $X$.  Because we have  selected only favorable hypersurfaces $X$, all divisors of $X$ are inherited from divisors of $V$.  In this work we will consider only effective divisors in $X$ that are inherited from \emph{effective} divisors in $V$, but more general effective divisors of $X$ are possible.  We thank M.~Stillman for explaining this point to us.}  The effective cone is not simplicial in general, so the ranges of the coefficients $a_i$ are not always obvious. However, one can consider non-negative linear combinations of the generators of $\lab{Eff}(X)$, which will by definition generate all effective divisors.

The Hodge numbers of the toric divisors $\check{D}^a$, which correspond to rays in the fan $F$ and therefore to points in $\Delta^\circ$, can be computed via polytope data alone~\cite{Braun:2016igl}, in the same fashion that the Hodge numbers of the Calabi-Yau are computed in~\cite{1993alg.geom.10003B, 0025-5726-29-2-A02}. We provide a brief summary of the results. As mentioned above, the polytopes $\Delta$ and $\Delta^\circ$ are dual, so there is a
one-to-one relation between faces of dimension $k$, $\Theta^{\circ [k]}$, of $\Delta^\circ$, and faces of dimension $3-k$,  $\Theta^{[3-k]}$, of $\Delta$.
The divisors $D^a$ can be organized according to their corresponding points in $\Delta^\circ$.  Let $l^*(\Theta)$ denote the number of interior points of a face $\Theta$; then:
\begin{itemize}
\item For divisors $D^a$ that correspond to vertices $\Theta^{\circ [0]}$ of $\Delta^\circ$, we have $h^{\bullet}(D, \mathcal{O}_D) = (1, 0, n)$, where $n= l^* \big(\Theta^{[3]}\big)$ and $\Theta^{[3]}$ is the three-dimensional face dual to $\Theta^{\circ [0]}$.
\item For divisors $D^a$ that correspond to points $v_a$ interior to one-dimensional faces $\Theta^{\circ [1]}$ of $\Delta^\circ$, we have $h^{\bullet}(D, \mathcal{O}_D) = (1, n, 0)$, where $n= l^*\big(\Theta^{[2]}\big)$ and $\Theta^{[2]}$ is the two-dimensional face dual to $\Theta^{\circ [1]}$.
\item For divisors $D^a$ that correspond to points $v_a$ that are interior to two-dimensional faces $\Theta^{\circ [2]}$ of $\Delta^\circ$, we have $h^{\bullet}(D, \mathcal{O}_D) = (n, 0, 0)$, where $n = l^*\big(\Theta^{[1]}\big) + 1$ and $\Theta^{[1]}$ is the one-dimensional face dual to $\Theta^{\circ [2]}$. If
    $n > 1$ then these divisors are reducible.
\end{itemize}
These facts make computing the Hodge numbers of toric divisors $\check{D}^a$ a simple combinatorial process. However, it is often the case that there are
fewer than $h^{1,1}$ linearly-independent rigid toric divisors, and therefore to search for instantons leading to a full rank $Q$ one must consider linear combinations of toric divisors that are not linearly equivalent to a toric divisor.  Because such combinations do not simply correspond to rays in the fan, obtaining their Hodge diamonds requires more effort. The Koszul sequence allows one to calculate this data, and has been implemented in the program {\tt cohomcalg}~\cite{Blumenhagen:2010pv,Blumenhagen:2010ed}, which we used extensively.  We refer the interested reader to~\cite{Blumenhagen:2010pv,Blumenhagen:2010ed} for details.

It is worth remarking that a linear combination of toric divisors that is rigid and irreducible is also necessarily singular.\footnote{We thank M.~Stillman and B.~Sung for helpful explanations of this point.} Consider a divisor $D$ that is linearly equivalent to $D_x + D_y$, where $D_x$ and $D_y$ are toric divisors defined by the vanishing of toric coordinates $x$ and $y$, respectively. In order for $D$ to be rigid we need $h^2(D)=0$, which implies that $h^2(D_x) = h^2(D_y) = 0$, as taking a linear combination will not affect the presence of these deformations. Then the only polynomial one can write to define the divisor is $xy=0$. This is singular along the intersection of the divisors $x = y = 0$. If the divisor is irreducible then $D_x$ and $D_y$ must have non-zero intersection, and therefore the point $x = y = 0$ is contained in the space, and $D$ is necessarily singular.  We find that of the 4390 triangulations in our ensemble that have a full-rank $Q$, 3422 remain full rank when only smooth toric divisors are included.

\begin{table}
\centering
\begin{tabular}{  l  c  c  c  }
 \hline
$h^{1,1}(X)$ & 2 & 3  & 4\\
\hline
 Number of polytopes & 36 & 244 & 1197 \\
Number of favorable polytopes & 36 & 243 & 1185 \\
 Number of favorable triangulations & 48 & 525 & 5330 \\
\hline
Number of full-rank triangulations & 24 & 262 & 4104 \\
\hline
Full-rank with only smooth divisors & 9 & 199 & 3214 \\
\hline
\end{tabular}
\caption{Results of the scan over reflexive polytopes with $h^{1,1}(X) \leq 4$.}\label{tab:scan}
\end{table}

\section{A Complete Scan at Small $h^{1,1}$} \label{scan}

Equipped with the results of \S\ref{sec1} and \S\ref{toric}, we computed the relevant topological data of all Calabi-Yau hypersurfaces in the Kreuzer-Skarke database with $2 \leq h^{1,1} \leq 4$.
We searched for divisors $D$ that are rigid linear combinations of up to three toric divisors $\check{D}$,
\begin{equation}
D = n_a \check{D}^a\,,
\end{equation}
where $n_a$ are nonnegative integers obeying $\underset{a}{\lab{max}}\,\, n_a =3$ and $\sum_a n_a\leq 3$.
We computed the topology of individual toric divisors via polytope data, and that of linear combinations with {\tt cohomcalg}. At $h^{1,1}=2,3,4$ we found that 24, 262, and 4104 triangulations, respectively, have full-rank $q$ matrices resulting from Euclidean D3-branes.
The results are summarized in Table~\ref{tab:scan}.\footnote{It sometimes happens that two isomorphic hypersurfaces are realized as hypersurfaces in different toric varieties corresponding to different polytopes. Since we are simply performing a scan over the geometries, we will not attempt to distinguish whether two Calabi-Yau hypersurfaces are different but will instead only refer to individual triangulations.}

\begin{figure}
	\centering
	\includegraphics[width=0.8\textwidth]{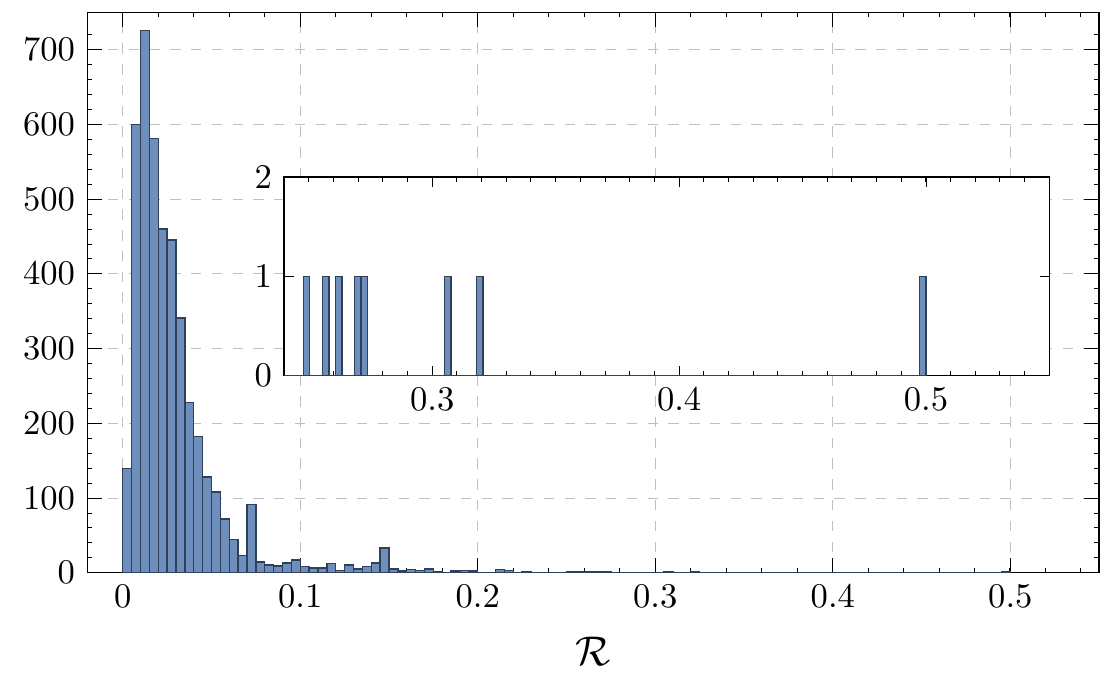}
	\caption{Histogram of geometric field ranges $\mathcal{R}$, in units of the reduced Planck mass $M_\lab{pl}$, for $h^{1,1} \leq 4$. The inset shows the tail of the distribution.}
	\label{fig:1}
\end{figure}

The combined field space radii for $h^{1,1} = 2,3,4$ are plotted in Figure~\ref{fig:1}. We find the maximum to be $\mathcal{R} \approx 0.5 \M$, in a case with $h^{1,1} = 3$, but in this example the overall volume of the Calabi-Yau is close to unity, and so the compactification is arguably not within the regime of perturbative control. The next largest is an example with $h^{1,1} = 4$ in which $\mathcal{R} \approx 0.3 \M$, and where the overall volume
is $\approx 20$.  This example is much better controlled, and therefore gives the upper bound that we report.

\begin{figure}
	\centering
	\includegraphics[width=0.8\textwidth]{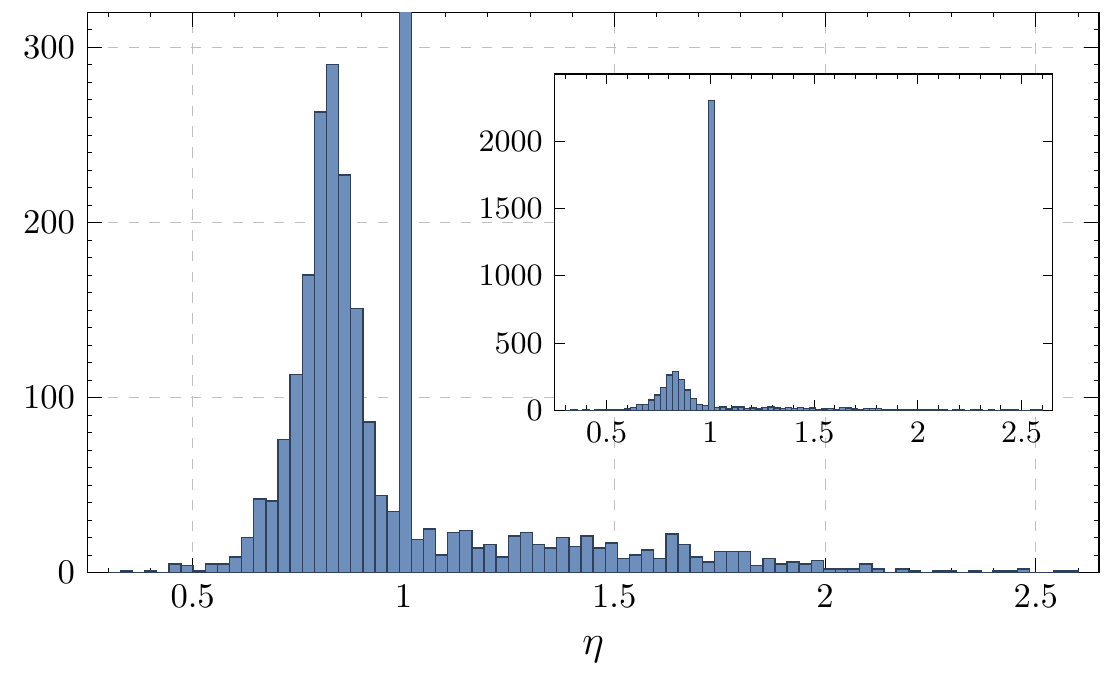}
	\caption{Histogram of enhancements $\eta$ for $h^{1,1} = 2, 3, 4$. Inset demonstrates the large peak at $\eta = 1$, i.e.~many geometries see no enhancement in size, or a reduction, from a non-trivial $Q$.}
	\label{fig:etaPlot}
\end{figure}

In Figure~\ref{fig:etaPlot}, we show a histogram of enhancements from lattice alignment, $\eta$, for the 4390 geometries with $\,h^{1,1} = 2, 3,$ and $4$. As seen in the inset, there is a spike at $\eta = 1$ corresponding to a large fraction of geometries---2180 out of 4390---that experience no enhancement from $Q$.  This occurs when the minimal generators of $\lab{Eff}(X)$ are rigid and thus the leading order $\mathcal{Q}$ is trivial.
In addition, many of the non-trivial $Q$-matrices actually decrease the geometric field range $\mathcal{R}$. We find a positive enhancement in 494 examples.

In this ensemble, the maximum enhancement from a nontrivial charge matrix is a factor of $\eta=2.6$ in a threefold with $h^{1,1} = 4$. The vertices of the polytope $\Delta^{\circ}$ are given by
\begin{align}
			\mathbf{d}_i = &\Big\{( 1, \minus 1, 0, 0),\, (\minus 1, 4, \minus 1, \minus 1), \,(\minus 1, \minus 1, 0, 0), \, (\minus 1, \minus 1, 1, 0), \\
			& (\minus 1, \minus 1, 0, 1), \,(\minus 1, 2, 0, 0), \,(\minus 1, \minus 1, 1, 1) \Big\}. \nonumber
		\end{align}

Here we have
\begin{equation}
\bm{\mathcal{Q}} = 2\pi \begin{pmatrix} ~~ 1 ~ & ~ 0 ~ & ~0 ~ & ~ 0~\\ ~\minus 1 ~ & \minus 1 ~ & ~1 ~ & ~1~ \\
~~0 ~ & ~0 ~ & ~1 ~ & ~0~ \\ ~~0 ~ & ~0 ~ & ~ 0 ~ & ~1~ \end{pmatrix}.
\end{equation}
This occurs in an example where the eigenvalues of $K_{ij}$ are quite small, and the geometric field range increases from $0.03 \M$ to only about $0.079 \M$. In this example not all of the rigid divisors are smooth. The next largest enhancement is $\eta=2.55$, which increases the geometric field range from $\mathcal{R} = .05$ to $\mathcal{R} = 0.12$. The vertices of the polytope $\Delta^{\circ}$ are given by
\begin{align}
			\mathbf{d}_i = &\Big\{( \minus 1, 2, \minus 1, \minus 1),\, (\minus 1, \minus 1, 2, 1), \,(\minus 1, \minus 1, 1, 1), \, (1, 0, \minus 1, \minus 1), \\
			& (\minus 1, \minus 1, 1, 2), \,(0, \minus 1, 1, 1), \,(2, 1, \minus 2, \minus 2) \Big\}, \nonumber
		\end{align}
and $Q$ is given by
\begin{equation}
\bm{\mathcal{Q}} = 2\pi \begin{pmatrix} ~~ 1 ~ & ~ 0 ~ & ~0 ~ & ~ 0~\\ ~~ 0 ~ & ~1 ~ & ~0 ~ & ~0~ \\
~~0 ~ & ~0 ~ & ~1 ~ & ~0~ \\ ~~2 ~ & ~2 ~ & ~ 1 ~ & \minus 1~ \end{pmatrix}.
\end{equation}
In this example all of the rigid divisors are smooth.

Although the scan at small $h^{1,1}$ did not yield a geometry that allows a parametrically large fundamental domain, some of the examples exhibit features that could be interesting for inflationary model building. Consider, for instance, the Calabi-Yau hypersurface in the toric variety $(\mathbbm{P}^1)^4$. The volume is
\begin{equation}
\mathcal{V} = 2\left(t_1 t_2 t_3 + t_1 t_2 t_4 + t_2 t_3 t_4 +  t_1 t_3 t_4\right),
\end{equation}
and the Mori cone conditions are simply $t_i > 0$, $i = 1,\dots, 4$. In this geometry, one can make the overall volume arbitrarily large while holding the largest eigenvalue of $K$ fixed, by taking $t_2 = t_3 = t_4 \equiv t_0$ for constant $t_0$, and letting $t_1 \equiv t \gg 1$. The largest eigenvalue of $K$ is then $1/(144t_0^4)\M^2$. This is an appealing feature, as suitably scaling up the volume can provide protection against some perturbative and nonperturbative corrections, while keeping the largest eigenvalue fixed at a sizable value.
For instance, by taking $t \rightarrow \infty$ and setting $t_0 = 0.2$, the largest eigenvalue of $K$ becomes $4.3\M^2$.
However, there is a\footnote{The remaining three divisors have large volumes for $t \rightarrow \infty$, and their contributions to the superpotential can be neglected.} divisor $D_s$ with volume $\tau_{s} = 6t_0^2 \approx 0.24$.  If there are higher-order instanton contributions\footnote{It is not clear that higher-order contributions from Euclidean D3-branes without flux will be nonvanishing, because $h^0 (kD) = h^2 (kD) = k$.
In fact, in this example are no rigid divisors at all: all of the toric divisors pulled back to the Calabi-Yau hypersurface have the Hodge numbers of K3 surfaces.
A superpotential might still be generated if worldvolume fluxes lift the zero modes corresponding to $h^2(D)$ deformations.} $\sim e^{-k\tau_{s}}$ for $k>1$, these are not necessarily negligible, e.g.~for $k=2$ their importance relative to the leading term is $e^{-2\pi(0.48)}/e^{-2\pi(0.24)} \sim 0.22$.

\section{Probing Large $h^{1,1}$} \label{largeHodge}

Our analysis thus far has been restricted to small Hodge numbers, $h^{1,1}\le 4$, but arguments in effective field theory and in random matrix theory suggest that new phenomena will appear in compactifications with $h^{1,1}\gg 1$ \cite{Bachlechner:2014gfa,toappearRMT}.   A comparative analysis of these proposals for alignment, and of the requisite degree of fine-tuning at the level of effective field theory, will appear in \cite{toappearRMT}; here we will briefly summarize the main ideas in order to provide orientation for our search at large $h^{1,1}$.

An influential early suggestion for alignment of $N \gg 1$ axions was the N-flation proposal \cite{Dimopoulos:2005ac}, where it was observed that the field range of a simple\footnote{The simplifying assumption is that $\mathcal{Q}=2\pi\mathbbm{1}$ in a basis in which $K_{ij}$ is diagonal.  This does not hold in generic examples, and in particular is violated in every geometry in our ensemble.} system of $N$ axions is the Pythagorean sum of the ranges of the individual axions, and schematically ${\cal R} \propto N^{1/2}$.  More recent works have identified stronger enhancements at large $N$.  Multi-axion alignment, the $N$-dimensional generalization of KNP alignment, yields exponentially large ranges, while plausibly requiring severe fine-tuning \cite{Choi:2014rja}.  Finally, in~\cite{Bachlechner:2014gfa} it was observed that generic charge matrices could give `spontaneous' field range enhancements as large as $N^{3/2}$ from a combination of lattice and kinetic alignment.  More precisely, the finding of~\cite{Bachlechner:2014gfa} is that for charge matrices $Q$ whose entries are well-approximated as independent and identically distributed (i.i.d.) variables, and are not too sparse, the distribution of field ranges takes the form
\begin{equation}
{\cal R} = N^{p} \zeta\,,
\end{equation} where $1\lesssim p \lesssim 3/2$ depends on the sparsity of $Q$.  Here $\zeta$ is a positive stochastic variable, varying from one realization of $Q$ to another, that has unit median and a {\it{heavy tail}} toward large values: in particular, the mean obeys $\langle \zeta \rangle \gg 1$.  The distribution of $\zeta$ is computable in special cases.  When the entries of $Q$ are such that $Q Q^{\top}$ is a Wishart matrix ${\cal{W}}$, one finds $\zeta \approx \lambda_1({{\cal W}})^{-1/2}$, with $\lambda_1({{\cal W}})$ the smallest eigenvalue of ${\cal{W}}$.
Because the probability density function of $\lambda_1({{\cal W}})$ has support near $\lambda_1=0$, $\zeta$ has a tail toward large positive values.
In turn, the range ${\cal R}$ has a heavy tail, and one expects to find, after a modest number of independent trials, a range ${\cal R}$ that exceeds the median value ${\cal R}_{\rm{med}}$ by orders of magnitude.

Both the engineered $N$-dimensional alignment of \cite{Choi:2014rja}, and the spontaneous alignment of~\cite{Bachlechner:2014gfa}, provide field-theoretic mechanisms for parametrically large field ranges.  However, it is clearly necessary to test these ideas in actual string compactifications, in order to understand whether quantum gravity indeed allows these effective theories, and so permits field ranges that are very large in Planck units.  To begin exploring this point, we will examine
a number of Calabi-Yau hypersurfaces, with $h^{1,1} \in \{50,60,70,80,90,100\}$.  More systematic results will appear in \cite{Khrulkov16}.

\subsection{Field ranges and volumes}
For ten geometries each at $h^{1,1} \in \{50, 60, 70, 80, 90, 100\}$, we computed the relevant topological and metric data and bounded the geometric field range $\mathcal{R}$.
Computing the topology of nontrivial linear combinations of toric divisors is computationally expensive at large $h^{1,1}$, so we only searched for rigid divisors among the toric divisors themselves.  In many cases the toric divisors suffice to lift all flat directions, and in such cases we bounded the field range. At large $h^{1,1}$, the vertex enumeration problem is computationally taxing and we used alternative methods to obtain the field range.

We may always trivialize $2N$ of the hyperplane constraints via the field transformation
	\begin{equation}
		\theta^i = \left(\mathcal{Q}^{-1}\right)\indices{^i_j} \Phi^j.
		\label{eq:xiBasis}
	\end{equation}
If $P = N$, this maps the fundamental domain $\mathcal{F}$ into the hypercube of side length $2\pi$. For $P > N$, the $2P$ hyperplane constraints
\begin{equation}
	\minus \pi \leq Q\indices{^a_i} \smash{\left(\mathcal{Q}^{-1}\right)\indices{^i_j}} \Phi^j \leq \pi
	\label{eq:addCuts}
\end{equation}
restrict $\mathcal{F}$ to a hypercube subject to $2(P-N)$ hyperplane `cuts.' In the $\Phi^i$ basis, distance in field space in Planck units is measured with respect to the metric
	\begin{equation}
		\bm{\Xi} = \left(\bm{\mathcal{Q}}^{-1}\right)^{\!\top} \!\!\cdot \mathbf{K} \!\cdot \!\bm{\mathcal{Q}}^{-1},
	\end{equation}
	whose maximum eigenvalue we denote $\xi_N^2$. If we temporarily ignore the additional constraints (\ref{eq:addCuts}), computing $\mathcal{R}$ via (\ref{eq:geomfieldrange}) involves evaluating the $\bm{\Xi}$-norm of $2^{N-1}$ vertices and is thus prohibitively expensive at large $N$.
However, we can always bound the geometric field range by
	\begin{equation}
		\mathcal{R} \le \mathcal{R}_{\rm{max}} = \pi \sqrt{N} \xi_N \M\,.
		\label{eq:cubeR}
	\end{equation}
At large $N$, eigenvector delocalization generally ensures that the ellipsoid's principal axes are nearly aligned with the diagonals of the hypercube, so that (\ref{eq:cubeR}) is often nearly saturated.  Upon including the additional $2(P-N)$ constraints (\ref{eq:addCuts}), the field range will be reduced by the maximally constraining cut, as detailed in~\cite{Bachlechner:2014gfa}. Because we always work with the full-rank square matrix $\mathcal{Q}\indices{^i_j} \subset Q\indices{^a_i}$, we approximate the field range using Eq. \ref{eq:cubeR}.
	
\begin{figure}
  \centering
  \includegraphics[width=0.5\textwidth]{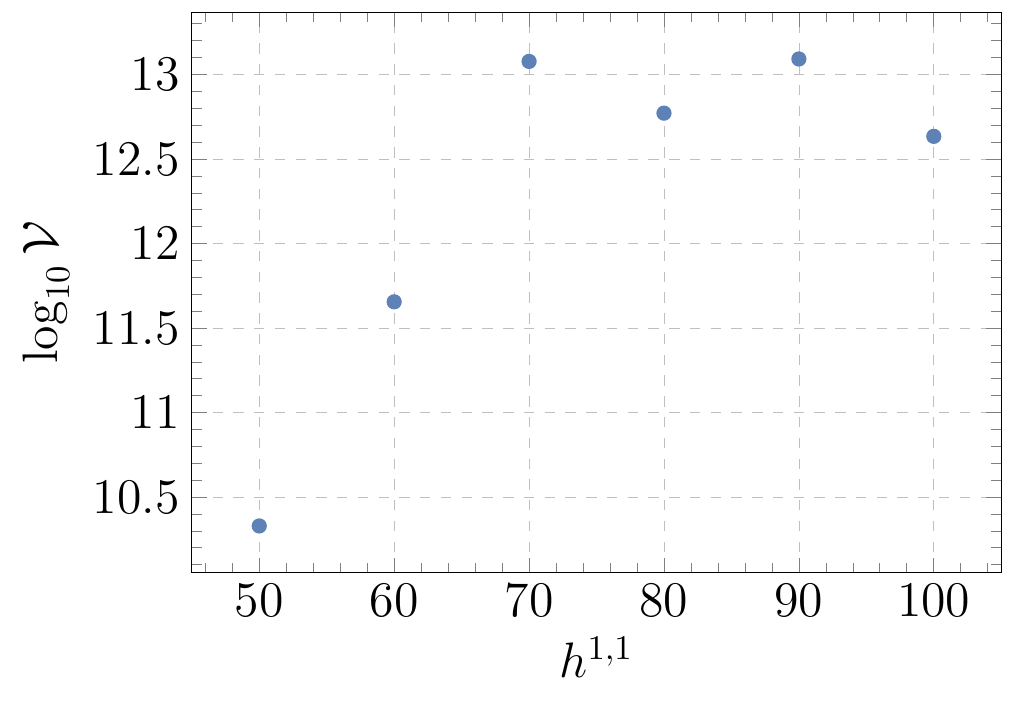}
  \caption{$\log_{10}$  of the mean volumes ${\cal{V}}$ as a function of $h^{1,1}$.}
  \label{fig:volums}
\end{figure}

\begin{figure}
	\begin{subfigure}[t]{0.48\textwidth}
		\centering
		\includegraphics[width=1\textwidth]{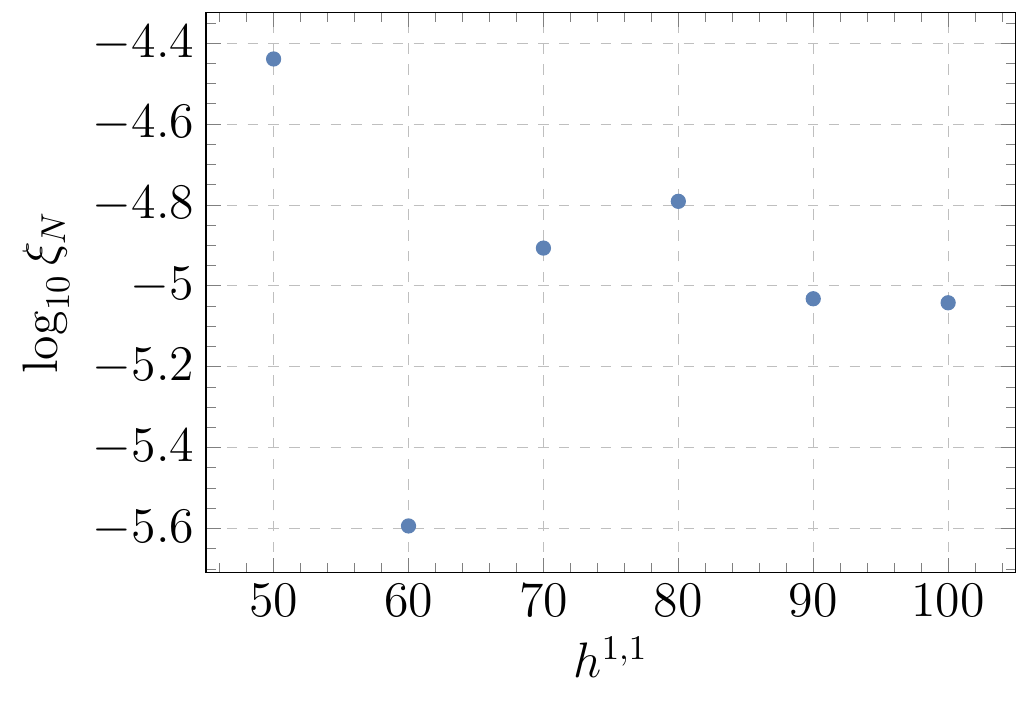}
		\caption{Average $\log_{10} \xi_N$ vs. $h^{1,1}$.}
		\label{fig:xiN}
	\end{subfigure}\hfill
	\begin{subfigure}[t]{0.48\textwidth}
		\centering
		\includegraphics[width=1\textwidth]{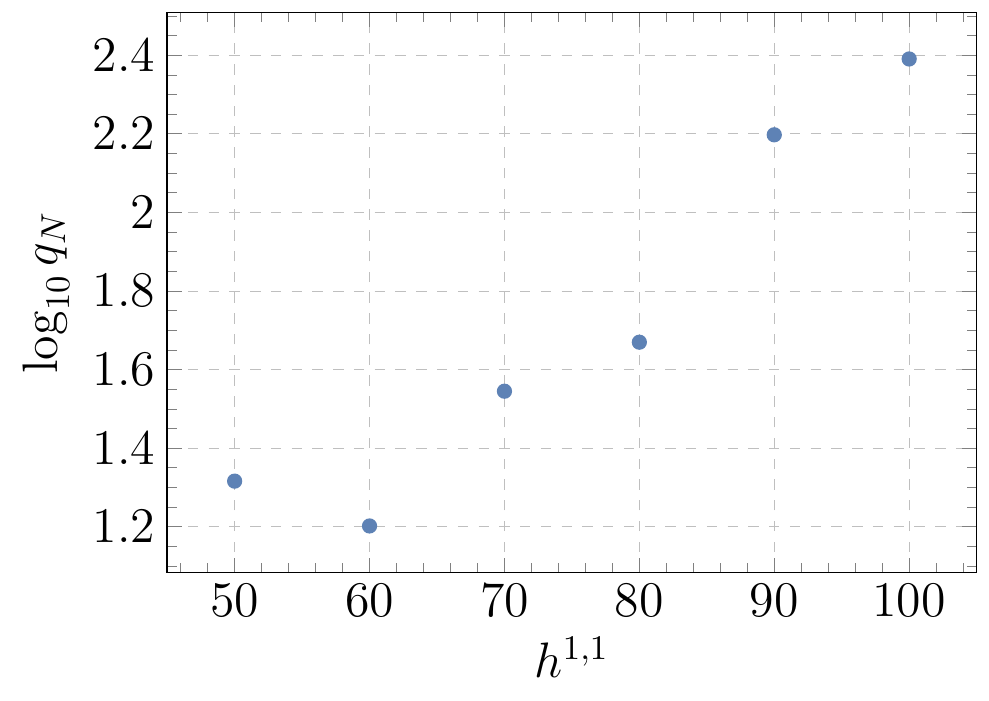}
		\caption{Average $\log_{10} q_N$ vs. $h^{1,1}$.}
		\label{fig:qq}
	\end{subfigure}~
	\caption{Average $\xi_N$ and $q_N$ at large $h^{1,1}$.}
\end{figure}

In all cases we find $\mathcal{R} \ll \M$. The mean volume of the Calabi-Yau at the apex of the stretched K\"ahler cone, as a function of $h^{1,1}$, is plotted in Figure~\ref{fig:volums}, and the mean value of $\xi_N$ as a function of $h^{1,1}$ is plotted in Figure~\ref{fig:xiN}. We also show $q_N$, the square root of the largest eigenvalue of $\left(\mathcal{Q}\mathcal{Q}^{\top}\right)^{\minus 1}$ in this basis, in Figure~\ref{fig:qq}.  We find the largest enhancement from lattice alignment occurs at $h^{1,1} = 100$, with $\eta_\lab{max} = 7.86$. We note that while the effect of alignment can be significant for $h^{1,1}\gg 1$, in our examples this is dwarfed by the growth of the volume with $h^{1,1}$. As $h^{1,1}$ grows, the number of holomorphic curves grows as well, giving more inequalities on the K\"ahler parameters to stay within the K\"ahler cone. By demanding that we remain in a regime of control, where all curve volumes are greater than one, the volume is forced to grow quite large (cf.~\cite{Denef:2004dm,Rudelius:2014wla}).

These characteristics are in stark contrast with those of the compactification studied by Denef et al.~in~\cite{Denef:2005mm}, where $h^{1,1} = 51$, but the volume was stabilized at $\mathcal{V} \sim 50$.  In~\cite{Denef:2005mm} the K\"ahler moduli were stabilized at a point where the smallest curve volumes were $0.2$, but even after scaling up the curve volumes to be $\geq 1$, one finds  $\mathcal{V} \sim 250$, which is vastly smaller than the volumes we find in hypersurfaces with comparable $h^{1,1}$.  A main reason that the volume can be kept small in~\cite{Denef:2005mm} is that the moduli space is very symmetric. The Calabi-Yau is constructed by taking identical toric patches and gluing them together, so the divisor and curve structure is simply repeated.
The result is that the overall volume of the Calabi-Yau does not increase dangerously with the curve volumes. In the two-parameter model of~\cite{Denef:2005mm}, denoting the volumes of the two classes of curves as $s$ and $u$, the overall volume takes the form $\mathcal{V} = s^3 +24 s^2 u+ 96 su^2 + 128u^3$, which is simple due to the symmetric intersection structure.

\subsection{The structure of $Q$}

We have seen that, although the largest eigenvalue $\left(\mathcal{Q}\mathcal{Q}^{\top}\right)^{\!\minus 1}$ was often quite large, the field range was still small.
Writing $\mathcal{Q}$ as
\begin{equation}
\mathcal{Q} = 2\pi\mathbbm{1} + \Delta_{\mathcal{Q}}\,,
\end{equation}
we expect (cf.~the analysis in \cite{Bachlechner:2014gfa,toappearRMT}) that if the entries of $\Delta_{\mathcal{Q}}$ are well-approximated as i.i.d.~stochastic variables, and if these entries are not too sparse,\footnote{Concretely, if e.g.~5\% of the entries of a $100\times 100$ matrix $\Delta_{\mathcal{Q}}$ are nonzero, the random matrix  analysis yields a heavy tail toward large ${\cal R}$.}  then ${\cal R}$ should manifest a large degree of enhancement from alignment.
In the geometries we examined, $Q$ is highly structured, and contains an identity matrix of size at least $h^{1,1}- 1$; the remainder $\Delta_{\mathcal{Q}}$ is then extremely sparse.\footnote{Between $1\%$ to $7\%$ of the entries in the large $h^{1,1}$ ensemble are populated, but the nonzero off-diagonal entries are restricted to a few rows and columns.}  We found that $\left(\mathcal{Q}\mathcal{Q}^{\top}\right)^{\!\minus 1}$ can in fact have a very large eigenvalue, but this is only necessary, not sufficient, for a large enhancement of the field range.  Indeed, we should interpret $q_N$ as the maximum possible enhancement from lattice alignment. The largest enhancement occurs when the largest-eigenvalue eigenvectors of the K\"{a}hler metric $K_{ij}$ and of $\left(\mathcal{Q} \mathcal{Q}^{\top}\right)^{\!\minus 1}$ are parallel, such that $\xi_{N}  = f_N q_N$, where $f_N^2$ is the largest eigenvalue of $K_{ij}$. If these eigenvectors are misaligned, the enhancement occurs in a different direction in field space---one that is irrelevant to the semi-diameter $\mathcal{R}$---and can compress the polytope, ultimately diminishing the field range.

Let us briefly  discuss why $Q$ so consistently contains a large identity matrix.
First consider a Calabi-Yau with large $h^{1,1}$ and small $h^{2,1}$. Here the large number of rigid divisors can be understood as a consequence of mirror symmetry.
If $h^{2,1}$ is small, then the number of points in the dual polytope $\Delta$ is small. Recall that the Hodge numbers of the toric divisors are computed by counting lattice points interior to faces of $\Delta$, so as $\Delta$ gets smaller the number of points interior to faces decreases, so more of the toric divisors have a better chance of becoming rigid.\footnote{We thank Andreas Braun for inspiration on this point.} For instance, consider the hypersurfaces in the Kreuzer-Skarke database with $h^{1,1} = 404$ and $h^{2,1} = 14$. There are six lattice polytopes corresponding to these Hodge numbers, and in all six at least 402 of the toric divisors are rigid.

On the other hand, this argument does not apply when both $h^{1,1}$ and $h^{2,1}$ are large. For example, we can consider a hypersurface with $h^{1,1} = h^{2,1} = 100$, whose corresponding $\Delta^\circ$ polytope has vertices	
		\begin{align}
			\mathbf{d}_i = &\Big\{( 1, \minus 1, \minus 1, \minus 1),\, (\minus 1, \minus 1, \minus 1, \minus 1), \,(\minus 1, \minus 1,  6, \minus 1), \, (\minus 1,  2, \minus 1, \minus 1), \\
			& (\minus 1, \minus 1,  6,  2), \,(\minus 1, \minus 1,  4,  5), \,(\minus 1,  2, \minus 1,  0), \,(\minus 1, \minus 1, \minus 1, 11), \,(\minus 1, \minus 1,  1,  9)\Big\}. \nonumber
		\end{align}
Here 98 of the 104 toric divisors have $h^{\bullet}(D,{\cal{O}}_D)=(1,0,0)$, even though the dual polytope has 134 points, only 8 of which are vertices. Therefore most of the dual cones have no interior points, and the non-vertex points are interior to only a few cones. This seems to be a consequence of the shape of $\Delta$, and is likely related to the requirement that the origin is the only interior point of $\Delta$: as the number of points included in the polytope grows, the shape must be more and more skewed.
In summary, we find it reasonable to conjecture that in many
geometries with large $h^{1,1}$, $Q$ will have a large-dimensional identity block, which does not contribute to lattice alignment.

\section{Conclusions} \label{sec:conclusions}

In this work we have initiated a systematic analysis of axion field ranges in type IIB compactifications on Calabi-Yau hypersurfaces in toric varieties. For axions descending from the Ramond-Ramond four-form $C_4$ in the 4390 geometries that we considered, we found a maximum field range of ${\cal R}_{\rm{max}}=0.3 \M$.
The largest enhancement of ${\cal R}$ due to lattice (KNP) alignment in our ensemble was a factor $2.6$, in an example with ${\cal R} \ll \M$. The numerical value of ${\cal R}_{\rm{max}}$ should not be overinterpreted, because
it can be made smaller or larger by imposing a more or less stringent requirement for control of the $\alpha^{\prime}$ expansion; the quoted value results from the requirement that the smallest curve has volume $(2\pi)^2\alpha'$.  What is clear is that in our examples, with our assumptions, the geometric field range does not parametrically exceed the Planck mass.

To assess the implications of these results, let us reexamine our assumptions and ask which of them might be relaxed.  First of all, it is plausible that in some geometries, one or more curves could be taken to have volume $t_i < 1$, while keeping other volumes large, without invalidating the sigma model expansion.  In this work we have followed a conservative, model-independent approach, but a more complete understanding of perturbative and nonperturbative corrections could allow for much larger field ranges.

Second, we considered axion potentials generated by Euclidean D3-branes wrapping divisors $D$ fulfilling the rigidity condition $h^{\bullet}(D,{\cal{O}}_D)=(1,0,0)$. That is, we required that $D$ be a rigid divisor of a smooth threefold, and did not incorporate the effects of orientifolding, worldvolume fluxes, bulk fluxes, and spacetime-filling seven-branes, which could alter the set of instanton contributions to the superpotential.
In particular, strong gauge dynamics on seven-branes, such as gaugino condensation on a stack of D7-branes coinciding with an O7-plane, provides a plausible mechanism for allowing larger field ranges, and more significant alignment, than we found in this work.  The axion periodicity induced by such branes is increased by a factor of the dual Coxeter number $c_2(G)$ of the condensing gauge group $G$, and many proposals for lattice alignment in string theory invoke stacks of D7-branes with $c_2>1$.  Systematically investigating such constructions would be valuable.

Third, we only examined $C_4$ axions in compactifications of type IIB string theory on Calabi-Yau hypersurfaces $X$ in toric varieties $V$, and we insisted that $X$ be favorable, meaning that all divisors of $X$ are inherited from $V$.  Each of these restrictions merits further investigation.  Two-form axions have a distinct parametric dependence on K\"ahler moduli, possibly allowing larger field ranges while maintaining control of the $\alpha^{\prime}$ expansion \cite{Grimm:2007hs,Rudelius:2014wla}.  We have no evidence to guide speculation about axion field ranges in threefolds that are not favorable hypersurfaces.

Finally, our systematic investigation occurred at small Hodge numbers, $h^{1,1} \le 4$, and we studied only a handful of examples with $h^{1,1}$ up to 100.
An analysis based on random matrix models, with parameters calibrated by the examples found here, suggests that the maximum field range at moderate $h^{1,1}$ could be large.  Whether this can occur in actual compactifications depends on a competition between a tendency for the overall volume $\cal{V}$ to grow with $h^{1,1}$, which suppresses the entries of the K\"ahler metric, and the fact that larger axion charge matrices $\mathcal{Q}$ can manifest a greater degree of lattice alignment.  We observed a tendency for $\mathcal{Q}$ to be close to the identity in cases with $h^{1,1} \gg 1$, which precludes large enhancements from alignment, due to the prevalence of rigid toric divisors in these examples.

In summary, in compactifications of type IIB string theory on Calabi-Yau hypersurfaces with $h^{1,1} \le 4$, Euclidean D3-branes wrapping divisors $D$ that do not intersect seven-branes give rise to a potential for $C_4$ axions that allows for a small degree of lattice alignment, which is insufficient to allow a super-Planckian geometric field range, in the absence of monodromy, in a parameter regime where all curves have volume $\ge (2\pi)^2\alpha'$.
Understanding the geometry of axion field space in far more general compactifications is an important problem for the future.

\section{Acknowledgements}
We thank A.~Braun and J.~Halverson for discussions, and thank V.~Khrulkov, M.~Stillman, and B.~Sung for collaboration on the related work \cite{Khrulkov16}.  We are particularly indebted to M.~Stillman for many helpful explanations.

\bibliographystyle{JHEP}
\bibliography{refs}
\end{document}